\documentclass{article}
\usepackage{t1enc}
\usepackage[utf8]{inputenc}
\usepackage[english]{babel}
\usepackage{graphicx}
\usepackage{amssymb}
\usepackage{amsmath}
\usepackage{amsthm}
\usepackage{bbm}

\def\OA{{\cal A}}
\def\DK{{\cal O}}
\def\SDK{{\cal K}}
\def\PR{{\cal P}}
\def\vNA{{\cal N}}
\def\Mink{{\cal M}}
\def\CO{{\bf C}}
\def\CO{{\bf C}}
\def\RE{{\bf R}}
\def\IN{{\bf Z}}
\def\NA{{\bf N}}
\def\fel{{\frac{1}{2}}}
\def\hfel{{\frac{3}{2}}}
\def\UN{\mathbbm{1}}
\def\qed{\ \vrule height 5pt width 5pt depth 0pt}
\def\cros{\raise1.9pt\hbox{$\scriptscriptstyle
          >$}\!\raise1.5pt\hbox{$\scriptstyle\triangleleft\,$}}
\theoremstyle{definition}\newtheorem{D}{Definition}
\theoremstyle{definition}
\theoremstyle{definition}\newtheorem{Prop}{Proposition}
\theoremstyle{definition}

\title{\bf Reichenbach's Common Cause Principle \\ in Algebraic Quantum Field Theory \\ with Locally Finite Degrees of Freedom}
\author{\textit{G\'abor Hofer-Szab\'o}\thanks{Department of Logic, E\"otv\"os University Budapest, email: gsz@szig.hu} \\
\textit{P\'eter Vecserny\'es}\thanks{Research Institute for Particle and Nuclear Physics, Budapest, email: vecser@rmki.kfki.hu}}
\date{ }

\begin{document}
\maketitle

\begin{abstract}
In the paper it will be shown that Reichenbach's Weak Common Cause Principle is \textit{not} valid in algebraic quantum field theory with locally finite degrees of freedom in general. Namely, for any pair of projections $A,B$ supported in spacelike separated double cones $\DK_a$ and $\DK_b$, respectively, a correlating state can be given for which there is no nontrivial common cause (system) located in the \textit{union} of the backward light cones of $\DK_a$ and $\DK_b$ and commuting with the both $A$ and $B$. 
Since noncommuting common cause solutions are presented in these states 
the abandonment of commutativity can modulate this result: noncommutative Common Cause Principles might survive in these models.
\vspace{0.1in}

\end{abstract}

\section{Introduction}

\textit{Reichenbach's Common Cause Principle} is the claim that if there is a correlation between two events $A$ and $B$ and there is no direct causal (or logical) connection between the correlating events then there always exists a common cause $C$ of the correlation. Reichenbach (1956) characterizes this common cause by four simple probabilistic conditions which together imply (and in this sense explain) the correlation in question.

Reichenbach's original definition of the common cause is formulated in a classical setting. Both the correlating events and the common cause are elements of a classical event algebra and the probability used for the correlation is a classical probability measure. This classical definition however can be generalized in a straightforward way to quantum events as well. In quantum theory events are typically represented by projections of a von Neumann algebra. These projections can be interpreted as 0-1--valued observables where the expectation value of a projection is the probability of the event that the observable takes on the value 1 in the appropriate quantum state. Hence, the Common Cause Principle can easily be translated into the non-classical case by simply replacing Boolean event algebras by projection lattices of von Neumann algebras and the classical probability measures by quantum states. 

If we consider a physical theory, classical or quantum, purely at algebraic-probabilistic level that is if we disregard the spatiotemporal location of and the dynamical connection between the events then the validity of the Common Cause Principle for the theory in question can readily be settled. One needs simply to check whether the probability measure space representing the theory probabilistically contains a common cause for every correlating (and causally not connected) pair of events. If it does then the Common Cause Principle can be said to hold for the theory in question; if it does not then the Principle fails to hold.

In this second case however one can have the following concern: What if our failure of finding a common cause for a given correlation is \textit{not} due to the fact that the common cause does not exist but it follows from the fact that the probabilistic description we apply to the physical situation in question is simply not ''fine'' enough to yield a detailed enough picture. In other words, the correlation \textit{does} have a common cause but it remains hidden behind our ''coarse'' description; a more detailed description of the same physical scenario however \textit{would} reveal this common cause. To show that such a concern can not be excluded \textit{a priori} we state here informally a proposition proved in (Hofer-Szab\'o, R\'edei, Szab\'o, 1999). Let $A$ and $B$ be two correlating events in a classical probability measure space. Then, there always exists an extension of this probability measure space such that in the extended space the correlation between $A$ and $B$ has a common cause in the Reichenbachian sense. Moreover, the proposition remains valid if we replace the classical probability measure space by a non-classical one. These propositions show that if we remain simply at the algebraic-probabilistic level we will never be able to falsify the Common Cause Principle. 
\vspace{0.1in}

\noindent
However, physical theories in general specify also the spatiotemporal location of events. The mathematically best elaborated theory localizing quantum events is algebraic quantum field theory or---using Haag's term---local quantum physics (Haag, 1992). In this theory observables (including projections as quantum events) are represented by $C^*$-algebras associated to open bounded regions of the given spacetime. To be specific, in this paper a $\mathcal{P}_\SDK$-covariant local quantum theory $\{\OA(\DK),\DK\in\SDK\}$ in a spacetime $\mathcal{S}$ with group $\mathcal{P}$ of geometric symmetries is meant:
\begin{itemize}
\item[(i)] A collection $\SDK$ of causally complete, bounded regions of $\mathcal{S}$, such that $(\SDK,\subseteq)$ is a net (i.e. a directed poset) under inclusion $\subseteq$.
\item[(ii)] An isoton map from $\SDK$ to unital $C^*$-algebras, $\SDK\ni\DK\mapsto\OA(\DK)$ which satisfies not only Einstein causality but also algebraic Haag duality: $\OA(\DK')'\cap\OA=\OA(\DK),\DK\in\SDK$. Here the primes mean spacelike complement and commutant, respectively. The so-called quasilocal observable algebra $\OA$ is defined to be the inductive limit $C^*$-algebra of the net $\{\OA(\DK),\DK\in\SDK\}$ of local $C^*$-algebras and $\OA(\DK')$ is the smallest $C^*$-algebra in $\OA$ containing the local algebras $\OA(\tilde\DK),\tilde\DK\subset\DK'$.
\item[(iii)] A group homomorphism $\alpha\colon\mathcal{P}_\SDK\to\textrm{Aut}\,\OA$ from the subgroup $\mathcal{P}_\SDK$ of $\mathcal{P}$ leaving the collection $\SDK$ invariant into the automorphism group of $\OA$ such that the automorphisms $\alpha_g,g\in\mathcal{P}_\SDK$ act covariantly on the net: $\alpha_g(\OA(\DK))=\OA(g\cdot\DK), \DK\in\SDK$.
\end{itemize}

The states of the local quantum system in question are defined to be states, i.e. normalized positive linear functionals on the quasilocal observable algebra $\OA$. One associates GNS representations $\pi\colon\OA\to\mathcal{B}(\mathcal{H})$ to these states and requires the existence of a unitary representation $U\colon\mathcal{P}_\SDK\to\mathcal{B}(\mathcal{H})$ that implements $\alpha$ in the representation $\pi$. One can examine the net of local von Neumann observable algebras given by weak clousures: $\vNA(\DK):=\pi(\OA(\DK))'', \DK\in\SDK$. Even one can start the local quantum theory in this von Neumann setting demanding further properties of the local von Neumann algebras and examining normal states of them.  A typical characteristics of Poincar\'e covariant theories is that for two projections $A$ and $B$ supported in spacelike separated regions there exists ''many'' normal states such that $A$ and $B$ correlate in that state. (For the precise sense of ''many'' see for example (Summers, Werner 1988) or (Halvorson, Clifton 2000).) In other words, correlations between spacelike separated events abound in local quantum theories. Since spacelike separation excludes direct causal connection we can only ask whether the Common Cause Principle holds in local quantum physics: Is it true that for every correlating projections $A$ and $B$ supported in spacelike separated regions there exists a projection $C$ among the local observables which satisfies the Reichenbachian criteria and hence can be interpreted as the common cause of the correlation? 

Since a local quantum theory does not allow free-floating observables one has to specify the above question a bit further and take a stand on where to localize the common cause $C$ of the correlation. A natural choice to place $C$ seems to be this: Consider the spacetime regions associated with the algebras containing the correlating events; take the intersection of the backward light cones of these regions; and place $C$ in a local algebra supported within this intersection.

The question whether the Common Cause Principle in the above characterized form is valid in a Poincar\'e covariant local quantum theory in the von Neumann setting was first raised by R\'edei (1997, 1998) and it proved to be an open question still now. Something weaker however could be decided. R\'edei and Summers (2002, 2005) have shown that if we localize the common cause not in the intersection but in the \textit{union} of the backward light cones of the supports of the local algebras then the Common Cause Principle can be verified. They called this version of the Principle the Weak Common Cause Principle. 

In this paper we intend to show that a crucial premise of the proof of R\'edei and Summers is that the algebras in question are \textit{von Neumann algebras of type III}. Although these algebras arise in a natural way in the context of Poincar\'e covariant theories other local quantum theories apply von Neumann algebras of other type. For example, theories with locally finite degrees of freedom are based on finite dimensional (type I) local von Neumann algebras. This raises the question whether the Weak Common Cause Principle is valid in general in local quantum theories? To give this question a more concrete form we choose a specific model here, the local quantum Ising model and investigate the status of the Common Cause Principle in this model. Our conclusion will be negative: even the Weak Common Cause Principle turns out to be \textit{not valid} in the Ising model in general. However, a simple example indicates that using noncommuting common cause systems the Common Cause Principle might be fulfilled.

In Section 2 we review the notion of the Reichenbachian common cause and its generalization to the situation where the correlation is due to a set of cooperating common causes, which is called a \textit{common cause system}. We also review the notion of the common cause (system) in the non-classical, i.e. in the quantum case and introduce a (possibly) noncommuting generalization. In Section 3 we briefly sketch the proof of the Weak Common Cause Principle in a Poincar\'e covariant local quantum theory given by R\'edei and Summers and point to the crucial premise concerning the type of the von Neumann algebra. In Section 4 we introduce the basics of the local quantum Ising model first. Then we construct faithful states in which none of the Common Cause Principles hold for commuting common cause systems. We close this section by presenting a simple example where noncommuting common cause systems do provide a solution even in these states. In Section 5 we conclude the paper.

\section{The notions of the common cause and common cause system in the classical and non-classical case}

Let $(\Sigma,p)$ be a classical probability measure space and let $A$ and $B$ be two positively correlating events in $\Sigma$ that is let 
\begin{equation} \label{corr}
p(A\wedge B)>p(A)\, p(B) .
\end{equation}
Reichenbach (1956) defines the common cause of the correlation (\ref{corr}) as follows:
\begin{D} \label{cc}
An event $C\in\Sigma$ is said to be the \textit{(Reichenbachian) common cause} of the correlation between events $A$ and $B$ if the following  
conditions hold: 
\begin{eqnarray}
p(A\wedge B|C)&=&p(A|C)p(B|C) \label{cc1}\\
p(A\wedge B|C^{\perp})&=&p(A|C^{\perp})p(B|C^{\perp}) \label{cc2}\\
p(A|C)&>&p(A|C^{\perp})  \label{cc3}\\
p(B|C)&>&p(B|C^{\perp}) \label{cc4}
\end{eqnarray}
where $C^{\perp}$ denotes the orthocomplement of $C$ and $p( \, \cdot \,|\, \cdot \,)$ is the conditional probability. Equations (\ref{cc1})-(\ref{cc2}) are called \textit{screening-off conditions}; inequalities (\ref{cc3})-(\ref{cc4}) are called \textit{positive statistical relevancy conditions}.
\end{D}
\vspace{0.1in}

\noindent
Reichenbach's original definition of the common cause, however, turned out to be too restrictive in two respects. First, it does not take into consideration those situations in which the correlation results from not a single cause but a \textit{system} of cooperating common causes. Second, to embrace common causes with ''negative'' impact on their effects, one has to drop the positive statistical relevancy conditions. As a result, a general common cause, or in other words, \textit{a common cause system} can be characterized simply as a factorizing partition in $\Sigma$:
\begin{D}\label{ccs}
A partition $\left\{ C_k \right\}_{k\in K}$ in $\Sigma$ is said to be the {\it common cause system} of the correlation between events $A$ and $B$ if the following screening-off condition holds for all $k\in K$: 
\begin{eqnarray} \label{ccs1}
p(A B\vert C_k)=p(A\vert C_k)\, p(B \vert C_k)
\end{eqnarray}
where $|K|$, the cardinality of $K$ is said to be the \textit{size} of the common cause system. A common cause system of size $2$ is called a common cause (without the adjective 'Reichenbachian', indicating that the inequalities (\ref{cc3})-(\ref{cc4}) are not required).
\end{D}
\vspace{0.1in}

\noindent
The notion of the common cause (system) can be generalized to the non-classical case. To do this, one replaces the classical probability measure space $(\Sigma,p)$ by the non-classical probability measure space $(\PR(\mathcal{N}), \phi)$ where $\PR(\vNA)$ is the (non-distributive) lattice of projections (events) of a von Neumann algebra $\vNA$. A finite set of mutually orthogonal projections $\left\{ C_k \right\}_{k\in K}\subset\mathcal{P}(\vNA)$ is called a \textit{partition of the unit} $\UN\in\vNA$ if $\sum_k C_k = \UN$. If $A,B \in \mathcal{P}(\mathcal{N})$ are two commuting projections which are (positively) correlated in the state $\phi\colon\mathcal{N}\to\CO$ that is
\begin{equation}\label{qcorr}
\phi(AB)>\phi(A)\, \phi(B)
\end{equation}
then the common cause (system) of this correlation is defined as follows:
\begin{D}\label{qccs}
A partition of the unit $\left\{ C_k \right\}_{k\in K}\subset\mathcal{P}(\mathcal{N})$ is said to be the {\em common cause system} of the correlation (\ref{qcorr}) between the commuting events $A, B\in\mathcal{P}(\mathcal{N})$ if for every $k\in K$
\begin{itemize}
\item[(i)] $C_k$ commutes with both $A$ and $B$; and
\item[(ii)] if $\phi(C_k)\not= 0$ then the following condition holds:
\end{itemize}
\begin{eqnarray}
\frac{\phi(ABC_k)}{\phi(C_k)}&=&\frac{\phi(AC_k)}{\phi(C_k)}
\frac{\phi(B C_k)}{\phi(C_k)}.
\label{qccs1} 
\end{eqnarray} 
A common cause system of size $\vert K\vert=2$ is called a {\em common cause}.
\end{D}
\vspace{0.1in}

\noindent
Using that $\UN=AB+A^{\perp}B^{\perp}+AB^{\perp}+A^{\perp}B$ the correlation (\ref{qcorr}) and the common cause condition (\ref{qccs1}) can be written as
\begin{eqnarray}
\phi(AB)\phi(A^{\perp}B^{\perp})&>&\phi(AB^{\perp})\phi(A^{\perp}B),
\label{qcorrrew}\\
\phi(ABC_k)\phi(A^{\perp}B^{\perp}C_k)&=&\phi(AB^{\perp}C_k)\phi(A^{\perp}BC_k).
\label{qccs1rew} 
\end{eqnarray} 
One can even allow here the case $\phi(C_k)=0$ since then both sides of (\ref{qccs1rew}) are zero. It is obvious from (\ref{qccs1rew}) that if $C_k\leq X$ with $X=A,A^\perp, B$ or $B^\perp$ for all $k\in K$ then $\left\{ C_k \right\}_{k\in K}$ serve as a common cause system independently of the chosen state $\phi$. These solutions are called \textit{trivial common cause systems}. In case of common cause, $\vert K\vert=2$, triviality means that $\{C_k\}=\{ A,A^\perp\}$ or $\{C_k\}=\{ B,B^\perp\}$. 

Definition \ref{qccs} requires the elements of the common cause system to commute with the events $A$ and $B$. This property plays an important role in lack of nontrivial solutions for common cause (systems) in the Ising model. Of course one could abandon this commutativity requirement, but then noncommutative conditionalization has to be allowed. If 
$\{ C_k \}_{k\in K}\subset\PR(\vNA)$ is a partition of the unit then the map
\begin{equation}\label{ncqcorr}
\vNA\ni A\mapsto E(A):=\sum_{k\in K} C_kAC_k\in\vNA
\end{equation}
defines a \textit{conditional expectation} $E\colon\vNA\to{\cal{C}}$ with ${\cal{C}}:=\sum_k C_k\vNA C_k$, i.e. $E$ is a unit preserving positive surjection onto the unital $C^*$-subalgebra  ${\cal{C}}\subseteq\vNA$ obeying the bimodule property $E(B_1AB_2)=B_1E(A)B_2; A\in\vNA, B_1, B_2\in{\cal{C}}$. We note that ${\cal{C}}$ contains exactly those elements of $\vNA$ that commute with 
$C_k,k\in K$.

\begin{D}\label{ncqccs}
A partition of the unit $\left\{ C_k \right\}_{k\in K}\subset\mathcal{P}(\mathcal{N})$ is said to be the (possibly) {\em noncommuting common cause system} of the commuting events $A,B\in\cal{P}(\vNA)$, which correlate in the state $\phi\colon\vNA\to\CO$, if the condition
\begin{eqnarray}
\frac{(\phi\circ E)(ABC_k)}{\phi(C_k)}&=&
\frac{(\phi\circ E)(AC_k)}{\phi(C_k)}
\frac{(\phi\circ E)(BC_k)}{\phi(C_k)}
\label{ncqccs1} 
\end{eqnarray} 
holds for those $k\in K$ when $\phi(C_k)\not= 0$. 
\end{D}
\vspace{0.1in}

\noindent
We note that if the events $A,B$ commute with $C_k, k\in K$ then $A,B\in{\cal{C}}$, therefore (\ref{ncqccs1}) leads to (\ref{qccs1}). Similarly to the commuting case, the requirements (\ref{ncqccs1}) can be written in an equivalent form:
\begin{equation}\label{ncqccsrew}
(\phi\circ E)(ABC_k))(\phi\circ E)(A^{\perp}B^{\perp}C_k)
=(\phi\circ E)(AB^{\perp}C_k)(\phi\circ E)(A^{\perp}BC_k),\ k\in K.
\end{equation}
With these definitions in hand now we turn to the analysis of the proof of R\'edei and Summers (2002) concerning the validity of the Weak Common Cause Principle in Poincar\'e covariant algebraic quantum field theory.

\section{The Weak Common Cause Principle in relativistic algebraic quantum field theories}

The basic data of a relativistic algebraic quantum field theory in the von Neumann setting is a net $\{\mathcal A (V)\}$ of local von Neumann algebras indexed by open, bounded subsets $V$ of the Minkowski space $\Mink$. In the paper of R\'edei and Summers (2002) this net of algebras is assumed to satisfy the more or less standard axioms such as (i) isotony; (ii) Einstein causality; (iii) relativistic covariance; (iv) irreducible vacuum representation; (v) weak additivity; (vi) type III local von Neumann algebras; (vii) local primitive causality. The axiom (vi), which can be derived from general principles (for example, from the existence of a scaling limit (Fredenhagen 1985)), is crucial, because  this axiom ensures the richness of projections in local algebras. Axiom (vii), local primitive causality means that for every nonempty convex region $V\subset\Mink$ the local algebras $\OA(V)$ and $\OA(V'')$ coincide, where $V''$ is the double spacelike complement, i.e. the causal completion of $V$. 

Let us consider \textit{a local system} $(\mathcal A(V_1),\mathcal A(V_2),\phi)$ where $V_1$ and $V_2$ are nonempty convex subsets in $\Mink$ such that $V''_1$ and $V''_2$ are spacelike separated double cones and $\phi$ is a locally normal and locally faithful state on the quasilocal algebra $\mathcal A$. Such states ''typically'' generate correlation between the projections $A \in\mathcal A(V_1)$ and $B\in\mathcal A(V_2)$. (See Summers, Werner 1988; Halvorson, Clifton 2000). However, by means of the axioms (i)-(v) one can prove that the algebras $\mathcal A(V_1)$ and $\mathcal A(V_2)$ of the local system are \textit{logically independent} (R\'edei 1995a, b). This raises the question whether the Common Cause Principle holds for the correlations of the local system.

To be more precise, one has to specify the localization of the possible common causes of the correlations. To this end, define the \textit{weak, common}, and \textit{strong past} of the regions $V_1$ and $V_2$, respectively, as:
\begin{eqnarray*}\label{wcspast}
wpast(V_1, V_2) &:=& I_-(V_1)\cup I_-(V_2) \\
cpast(V_1, V_2) &:=& I_-(V_1)\cap I_-(V_2) \\
spast(V_1, V_2) &:=& \cap_{x \in V_1 \cup V_2}\, I_-(x),
\end{eqnarray*}
where $I_-(V)$ denotes the union of the backward light cones $I_-(x)$ of every point $x$ in $V$. (In (R\'edei, Summers 2005) the regions $I_-(V_i)\setminus V_i, i=1,2$ are used in the definition of \textit{wpast} and \textit{cpast}. However, if assumption (vii), i.e. local primitive causality holds the corresponding algebras coincide.) Then a local system $(\mathcal A(V_1),\mathcal A(V_2),\phi)$ is said to \textit{satisfy the Weak Common Cause Principle} if for any pair  $A \in\mathcal A(V_1)$, $B\in\mathcal A(V_2)$ of correlating projections  there exists a finite set $\{ C_k \}_{k\in K}$ of projections in the local von Neumann algebra $\mathcal A(V)$  associated with a bounded region $V\subset\Mink$ such that 
\begin{description}
\item[(i)] $\{ C_k \}_{k\in K}$ is a \textit{nontrivial} common cause system of the correlation (\ref{qcorr}) in the sense of Definition \ref{qccs},
\item[(ii)] $V \subset wpast(V_1, V_2)$.
\end{description}
Replacing $wpast(V_1, V_2)$ by $cpast(V_1, V_2)$ ($spast(V_1, V_2)$) in (ii) one obtains the \textit{(Strong) Common Cause Principle}. Replacing the commuting common cause system by a \textit{noncommuting} one (that is Definition \ref{qccs} by Definition \ref{ncqccs} in (i)) we speak about \textit{noncommutative (Weak, Strong) Common Cause Principle}. This noncommutative version will be shortly discussed in the next Section.

Proposition 3 of the paper of R\'edei and Summers states that if the net $\{\mathcal A(V)\}$ satisfies conditions (i)-(vii) then every local system $(\mathcal A(V_1),\mathcal A(V_2),\phi)$ satisfies the Weak Common Cause Principle (with $\vert K\vert =2$). Therefore algebraic quantum field theories with properties (i)-(vii) give rise not only to correlations between spacelike separated regions but also to a common causal explanation of the correlations. Whether this Proposition remains valid if weak past is substituted by common or strong\footnote{If one admits \textit{un}bounded regions as well then the Strong Common Cause Principle is known to be false: von Neumann algebras pertaining to complementary wedges contain correlated projections but the strong past of such wedges is empty (see Summers and Werner, 1988 and Summers, 1990).} past is an open question. We are of the opinion that the proof of the validity of the (Strong) Common Cause Principle would require the knowledge of an explicit dynamics. Namely, there is no bounded region $V$ in $cpast(V_1, V_2)$ (hence neither in $spast(V_1, V_2)$) for which isotony would ensure that $\OA(V_1\cup V_2)\subset\OA(V'')$. But dynamics relates the local algebras: $\OA(V_1\cup V_2)\subset\OA(V''+t)=\alpha_t(\OA(V''))$ can be fulfilled for certain $V\in cpast(V_1, V_2)$ and for certain time translation by $t$. 

The proof of the mentioned proposition of R\'edei and Summers is based on two Lemmas (Lemma 3 and 4). The first Lemma gives a sufficient condition for a projection $C$ to be a nontrivial common cause ($C:= C_1, \vert K\vert=2$) of projections $A,B$ correlating in a faithful state $\phi$ of the von Neumann algebra $\vNA$. If one supposes that $C<AB$ then $\{ C,C^\perp\}$ will be a nontrivial solution: (\ref{qccs1rew}) fulfills trivially for $C$ (both sides are zero) and the value $\phi(C)$ is determined from (\ref{qccs1rew}) for $C^\perp:=C_2$. It turns out that 

$$0< \phi(C)= \frac{\phi(AB)\phi(A^\perp B^\perp)-\phi(AB^\perp)\phi(A^\perp B)}{\phi(A^\perp B^\perp)} < \phi(AB).$$

The second Lemma states that in a type III von Neumannn algebra $\vNA$ with a faithful normal state $\phi$ for every projection $P\in\PR(\vNA)$ and every positive real number $0< r< \phi(P)$ there exists a projection $C\in\PR(\vNA)$ such that $C<P$ and $\phi(C)=r$. 

The properties of the local quantum theory in question ensure that both Lemmas are applicable and one can localize $C$ in a bounded region $V\subset wpast(V_1, V_2)$ for which $V_1\cup V_2\subset V''$. It is so because $\OA(V_1\cup V_2)\subset \OA(V)$ by isotony and local primitive causality, hence $A,B$ and $C$ all reside in the common local von Neumann algebra $\OA(V)$, which is of type III due to (vi).  

The assumption that the von Neumann algebra $\mathcal N$ is of type III is crucial in the second Lemma. This allows to find a subprojection $C$ of $P=AB$ with a prescribed value $r=\phi(C)<\phi(P)$ coming from the common cause system condition (\ref{qccs1}) for $C^\perp$. Although type III local von Neumann algebras are natural in the context of Poincar\'e covariant algebraic quantum field theories other local theories apply other type of von Neumann algebras, which do not obey a similar richness of commuting projections. The next Section shows that there is a tight connection between the fate of the commutative Common Cause Principle and the type of the local algebras. 

\section{The Common Cause Principle in the Ising model}

\begin{figure}[ht]
\centerline{\resizebox{7cm}{!}{\includegraphics{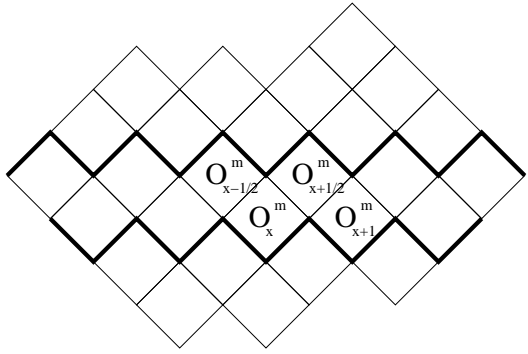}}}
\caption{A thickened Cauchy surface in the two dimensional Minkowski space $\Mink^2$}
\label{cauchy}
\end{figure}

Quantum spin chains can be formulated in terms of algebraic quantum field theory. Here we give a brief overview of how our concrete model fits in the general settings of algebraic quantum field theory. The local observable algebras of $G$-spin chains (Szlach\'anyi, Vecserny\'es, 1993) or Hopf spin chains (Nill, Szlach\'anyi, 1997) are labelled by the net of `intervals' $(i,j):=\{i,i+\fel,\dots,j-\fel,j\}\subset\fel\IN$ of half-integers. The `one-point' observable algebras $\OA(i,i), i\in\fel\IN$ of a Hopf spin chain are isomorphic to the finite dimensional Hopf $C^*$-algebra $H$ or its dual $\hat H$, i.e. there are local isomorphisms $A_i\colon H\to\OA(i,i), i\in\IN$ and $A_i\colon\hat H\to\OA(i,i), i\in\IN+\fel$. In case of $G$-spin chains $H$ is the group algebra $\CO G$ of a finite group $G$ and $\hat H$ is $\CO(G)$, the algebra of complex functions on $G$. If $G=\IN_2:=\{ e,g\vert g^2=e\}$, the cyclic group of order 2, one gets the Ising quantum chain. The local observable algebra $\OA(i,j)$ corresponding to an interval $(i,j)$ is defined to be the iterated crossed product $A(i,i)\cros\OA(i+\fel,i+\fel)\cros\dots\cros\OA(j,j)$ of dually related Hopf algebras with respect to the left Sweedler action on their dual. The set of half-integers can be interpreted as the space coordinates of the center $(0,x)$ and $(1/2,x+1/2), x\in\IN$ of minimal double cones $\DK^m_x$ of unit diameter on a thickened Cauchy surface in two dimensional Minkowski space $\Mink^2$ (M\"uller, Vecserny\'es, to be published). (See Fig. \ref{cauchy}.) An interval $(i,j)\subset\fel\IN$ can be interpreted as the smallest double cone $\DK_{i,j}\subset\Mink^2$ containing both $\DK_i^m$ and $\DK_j^m$. They determine a directed subset $\SDK^m_{CS}$ of double cones in $\Mink^2$, which is left invariant by the group of space-translations with integer values, that is $\PR_{\SDK^m_{CS}}=\IN$. Identifying the local algebras with those of the spin chain, $\OA(\DK_{i,j}):=\OA(i,j)$ one gets a $\PR_{\SDK^m_{CS}}$-covariant local quantum theory in $\Mink^2$ described in Chapter 1: The quasilocal observable algebra is isomorphic to the two-sided infinitely iterated crossed product: $\OA\simeq\ldots\cros H\cros\hat H\cros H\cros\hat H\cros\ldots$. Isotony, locality, algebraic Haag duality, and discrete space-translation covariance, i.e. $\IN$-covariance, hold in these models. We note that for $n\in\NA$ the local algebras $\OA(\DK_{i,i-\fel+n}), i\in\fel\IN$ are isomorphic to the full matrix algebra $M_{{\vert H\vert}^n}(\CO)$, where $\vert H\vert$ is the dimension of the underlying group algebra (or Hopf algebra) $H$. Hence, the quasilocal observable algebra is a uniformly hyperfinite (UHF) $C^*$-algebra of type $\vert H\vert^\infty$. 

In case of the Ising model the causal (integer valued) time evolutions are classified (M\"uller, Vecserny\'es, to be published). That is the possible causal and (discrete) time translation covariant extensions of the ''Cauchy surface net'' 
$\{ \OA(\DK),\DK\in\SDK^m_{CS}\}$ to $\{\OA(\DK),\DK\in\SDK^m\}$ are given, where $\SDK^m$ is the subset of double cones in $\Mink^2$ that are spanned by minimal double cones of unit diameter being integer time translates of those in $\SDK^m_{CS}$, i.e. those in the original Cauchy surface. The set $\SDK^m$ is left invariant by integer space and time translations, that is $\PR_{\SDK^m}=\IN\times\IN$, and the extended net also satisfies isotony, locality and algebraic Haag duality. Moreover, the commuting (unit) time and (unit) space translation automorphisms $\beta$ and $\alpha$ of the quasilocal algebra $\OA$ act covariantly on the local algebras. The causal time translation automorphisms $\beta$ of $\OA$ can be parametrized by $\theta_1,\theta_2;\eta_1,\eta_2$ with $-\pi/2 <\theta_1,\theta_2\leq\pi/2$ and $\eta_1,\eta_2\in\{1,-1\}$ and they are given on an algebraic generator set $\{U_i\in\OA(\DK^m_i), i\in\fel\IN\}$ of $\OA$. In terms of the local embeddings of $\CO\IN_2$ 
 and $\CO(\IN_2)$ into $\OA$
\begin{eqnarray}\label{gen_set}
U_i:= \left\{ \begin{array}{rl} A_i(g), & i\in\IN, 
\\ A_i(\chi_e-\chi_g), & i\in\IN+\fel ,\end{array}\right.
\end{eqnarray}
where the $\chi_e,\chi_g\in\CO(\IN_2)$ are the characteristic functions of the corresponding group element. The generators are selfadjoint unitaries satisfying the (commutation) relations
\begin{eqnarray}\label{comm_rel}
U_i U_j = \left\{ \begin{array}{rl} -U_j U_i, & \mbox{if}\ |i-j|=\fel
\\ U_j U_i, & \mbox{otherwise}\ \end{array} \right.
\end{eqnarray}
due to the iterated crossed product structure of the chain. The  automorphisms $\beta=\beta(\theta_1,\theta_2,\eta_1,\eta_2)$ of $\OA$ corresponding to causal time translations by a unit read as
\begin{eqnarray}\label{causal_automorph1}
\beta(U_x)&=&\eta_1 \sin^2\theta_1 U_x+\eta_1 
      \cos^2\theta_1 U_{x-\fel}U_xU_{x+\fel}\nonumber\\
  &&\quad+\frac{i}{2}\sin2\theta_1(U_{x-\fel}U_x-U_xU_{x+\fel}),\\
\label{causal_automorph2}
\beta(U_{x+\fel})&=&\eta_2\sin^2\theta_2 U_{x+\fel} 
         +\eta_2\cos^2\theta_2\beta(U_x)U_{x+\fel}
         \beta(U_{x+1})\nonumber\\
 &&\quad+\frac{i}{2}\sin2\theta_2(\beta(U_x)U_{x+\fel}-
  U_{x+\fel}\beta(U_{x+1})),
\end{eqnarray}
where $x\in\IN$. The causal evolutions clearly show how local primitive causality holds in the discrete case: If $V$ consists of three neighbouring minimal double cones on a thickened Cauchy surface then $\OA(V)=\OA(V'')$ due to (\ref{causal_automorph1}) and (\ref{causal_automorph2}).

\begin{figure}[ht]
\centerline{\resizebox{10cm}{!}{\includegraphics{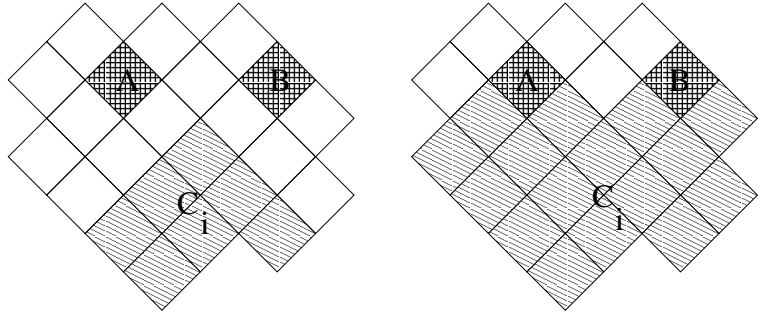}}}
\caption{The Common and the Weak Common Cause Principle}
\label{strongandweak}
\end{figure}

Fixing a causal time evolution let us choose two projections $A\in\OA(\DK_a)$ and $B\in\OA(\DK_b)$ in two local algebras corresponding to spacelike separated double cones $\DK_a,\DK_b\in\SDK^m$. Giving a state on $\OA$ that leads to positive correlation for $A$ and $B$ one can ask if there is a nontrivial common cause (system) in a local algebra $\OA(\DK)$, where $\DK\in\SDK^m$ is contained in the union of backward cones of $\DK_a$ and $\DK_b$. If the answer is affirmative, i.e. the Weak Common Cause Principle holds, one can look for the common cause (system) having the ''smallest'' localization region and can check whether it satisfies the (Strong) Common Cause Principle or not (see Fig. \ref{strongandweak}). We note that in our case bounded regions of $\Mink^2$ mean finite unions of minimal double cones $\DK^m$ and the definition of strong past is modified as $spast(V_1,V_2):=\cap_{\DK^m\subset V_1\cup V_2} I_-(\DK^m)$. 

\begin{Prop}\label{failWCCP}
Fixing a causal time evolution let us choose two nonzero projections $A\in\OA(\DK_a)$ and $B\in\OA(\DK_b)$ localized in two spacelike separated double cones $\DK_a,\DK_b\in\SDK^m$. One can construct faithful states on $\OA$ such that the Weak Common Cause Principle fails. 
\end{Prop}

\noindent\textit{Proof.} Since $\OA$ is a UHF algebra there is a unique (non-degenerate) normalized trace $\textrm{Tr}\colon\OA\to\CO$ on it, which coincides with the unique normalized trace on any unital full matrix subalgebras of $\OA$. As we have already mentioned local algebras corresponding to double cones $\DK_{i,i-\fel+n}\in\SDK^m_{CS}$ with $i\in\fel\IN, n\in\NA$ are isomorphic to full matrix algebras $M_{2^n}(\CO)$. One can find double cones $\tilde\DK_x\supseteq\DK_x, x=a,b$ in $\SDK^m$ that are also spacelike separated and are (integer) time translates of cones $\DK_{i(x),i(x)-\fel+n(x)}\in\SDK^m_{CS}$ with $i(x)\in\fel\IN$ and $n(x)\in\NA$ for $x=a,b$. Then $\OA(\tilde\DK_x)$ is isomorphic to the full matrix algebra $M_{2^{n(x)}}(\CO)$ due to time translation covariance. Let $\tilde\DK\in\SDK^m$ be a double cone that contains both $\tilde\DK_a$ and $\tilde\DK_b$ and that is a time translate of a cone $\DK_{i,i-\fel+n}\in\SDK^m_{CS}$ with $i\in\fel\IN$ and $n\in\NA$. Therefore $\OA(\tilde\DK)$ is isomorphic to the full matrix algebra $M_{2^n}(\CO)$. Hence, $A,A^{\perp}$ and $B,B^{\perp}$ are projections in two commuting full matrix algebras in a full matrix algebra, that is the (mutually orthogonal) projections  $P=AB,A^{\perp}B^{\perp},AB^{\perp},A^{\perp}B$ have nonzero rational traces $m_P/2^n$ with $m_P\in\NA$ and $\sum_P m_P=2^n$. Then  
\begin{equation}\label{failstate}
X \mapsto \phi(X)_\lambda:=\textrm{Tr} (\sum_P\lambda_P
\frac{2^n}{m_P}P X), \quad 0< \lambda_P,\ \sum_P\lambda_P=1
\end{equation}
defines a faithful state $\phi_\lambda$ on $\OA$ due to the faithfulness of the trace. The requirement of positive correlation $\phi_\lambda(AB)>\phi_\lambda(A)\phi_\lambda(B)$ and the common cause equation (\ref{qccs1rew}) read as
\begin{eqnarray}\label{pcandcc1} 
\lambda_{AB}\lambda_{A^{\perp}B^{\perp}} &>& \lambda_{AB^{\perp}}\lambda_{A^{\perp}B},\\
\frac{\lambda_{AB}\lambda_{A^{\perp}B^{\perp}}}
{ m_{AB}m_{A^{\perp}B^{\perp}}}
 {\textrm{Tr}}(ABC_k){\textrm{Tr}}(A^{\perp}B^{\perp}C_k)
&=&\frac{\lambda_{AB^{\perp}}\lambda_{A^{\perp}B}}
{m_{AB^{\perp}}m_{A^{\perp}B}}
{\textrm{Tr}}(AB^{\perp}C_k){\textrm{Tr}}(A^{\perp}BC_k).\label{pcandcc2} 
\end{eqnarray}
Let us choose the $\lambda$ parameters in a way to satisfy  (\ref{pcandcc1}), moreover let the products $\lambda_{AB}\lambda_{A^{\perp}B^{\perp}}$ and $\lambda_{AB^{\perp}}\lambda_{A^{\perp}B}$ be rational and irrational, respectively. Such numbers trivially exist; e.g. $\lambda_{AB}=\lambda_{A^{\perp}B^{\perp}}= \frac{1}{4}$, $\lambda_{AB^{\perp}}=\frac{1}{4}+\frac{\pi}{20}$ and $\lambda_{A^{\perp}B}= \frac{1}{4}-\frac{\pi}{20}$. However, if the projections $C_k, k\in K$ are elements of a local (hence, finite dimensional) algebra $\OA(\DK_c)$ then the traces of the products of commuting projections in (\ref{pcandcc2}) have rational values. Thus  (\ref{pcandcc2}) fulfills only if both sides are zero, that is only if $C_k\leq X$, with $X=A, A^{\perp}, B, B^{\perp}$ for $k\in K$. Therefore all of the solutions are trivial common cause systems\footnote{The weak closure $\vNA$ of the image $\pi_\lambda(\OA)$ in the GNS representation $\pi_\lambda$ corresponding to $\phi_\lambda$ is a type II hyperfinite factor (see. e.g. (Murphy, 1990)). Therefore nontrivial quasilocal common causes may exist in $\vNA$ since they can have irrational traces.}, which are excluded by definition in the Weak Common Cause Principle.\qed

Obviously, the Weak Common Cause Principle is weaker then the (Strong) Common Cause Principle, hence the Proposition above falsifies the latter principles as well. The counter-example in the proof is purely an algebraic-probabilistic one, there has been no mention of the localization of the projections $C_k$. The crucial property is that $C_k$ should belong to a local algebra. But local algebras are finite dimensional $C^*$-algebras in the Ising model (which coincide in the GNS-representation with their weak closures, which are type I Neumann algebras then), and the normalized trace of commuting projections in such a local algebra is a rational number. 

Our proof can be trivally extended to Hopf spin models with causal dynamics (or in general to UHF  quasilocal observable algebras). The only difference is that the simple local algebras are isomorphic to $M_{\vert H\vert^n}(\CO),n\in\NA$, where the dimension $\vert H\vert$ of the underlying Hopf algebra $H$, which is equal to 2 only in the Ising model.
\vspace{0.2in}

\noindent
Having shown that the commutative Weak Common Cause Principle is not universally valid in local quantum theories one can ask whether the noncommutative (Weak, Strong) Common  Cause Principle remains valid. Here we will not approach the full problem but only show that the state (\ref{failstate}) which falsified the Weak Common Cause Principle in the Ising model is not necessarily a falsifying state for the \textit{noncommutative} Common Cause Principle.

To this end, let us fix two parameters of the causal dynamics $\beta$ in (\ref{causal_automorph1}) : $\theta_1=0,\eta_1=1$. Let us choose two projections with spacelike separated supports:
\begin{eqnarray}\label{ABprojections}
A&:=&\fel\beta(\UN+U_0)=\fel(\UN+U_{-\fel}U_0U_\fel)\in\OA(\DK^m(1,0)),\\
B&:=&\fel\beta(\UN+U_1)=\fel(\UN+U_{\fel}U_1U_\hfel)\in\OA(\DK^m(1,1)),
\end{eqnarray}
where $\DK^m(t,x)\in\SDK^m$ is a minimal double cone centered in time and space coordinates $t$ and $x$, respectively. Let us choose the state $\phi_\lambda$ of $\OA$ given in (\ref{failstate}) with the specific values mentioned in Prop. \ref{failWCCP}: $\lambda_{AB}=\lambda_{A^{\perp}B^{\perp}}= \frac{1}{4}$, $\lambda_{AB^{\perp}}=\frac{1}{4}+\frac{\pi}{20}$ and $\lambda_{A^{\perp}B}= \frac{1}{4}-\frac{\pi}{20}$. Hence, $A$ and $B$ are positively correlated in $\phi_\lambda$. Using that the trace of any monomial of the generators $U_i,i\in\fel\IN$ is zero a straightforward calculation shows that for the projection $C\in\OA(\DK_{0,1})$ with $\DK_{0,1}\in\SDK^m_{CS}\subset\SDK^m$ given by 
\begin{equation}\label{NCCCforAB}
C=\fel(\UN+a_1U_\fel+a_2U_1+ia_3U_0U_\fel);
\quad a_1,a_2,a_3\in\RE,\ \sum_{i=1}^3 a_i^2=1,
\end{equation}
the following equations hold:
\begin{eqnarray}\label{fik}
(\phi \circ E) (ABC) & = & \frac{1}{16} \big( \lambda_{AB} +  \lambda_{A^{\perp}B^{\perp}} a_1^2 + \lambda_{A^{\perp}B} (a_2^2 + a_3^2)\big), \\
(\phi \circ E) (A^{\perp}B^{\perp}C) & = & \frac{1}{16} \big( \lambda_{AB} a_1^2  +  \lambda_{A^{\perp}B^{\perp}} + \lambda_{AB^{\perp}} (a_2^2 + a_3^2)\big), \\
(\phi \circ E) (AB^{\perp}C) & = & \frac{1}{16} \big(\lambda_{A^{\perp}B^{\perp}} (a_2^2 + a_3^2) + \lambda_{AB^{\perp}} +  \lambda_{A^{\perp}B} a_1^2 \big), \\
(\phi \circ E) (A^{\perp}BC) & = & \frac{1}{16} \big( \lambda_{AB} (a_2^2 + a_3^2) +  \lambda_{AB^{\perp}} a_1^2 + \lambda_{A^{\perp}B} \big).
\end{eqnarray}

\begin{figure}[ht]
\centerline{\resizebox{3cm}{!}{\includegraphics{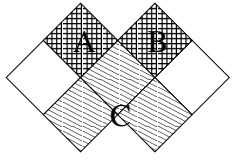}}}
\caption{Localization of the noncommuting common cause of the correlation}
\label{strong}
\end{figure}

Hence, (\ref{ncqccs1}) fulfil for $C$ and also for $C^\perp$ because the equations above are quadratic in the parameters $a_1,a_2,a_3$. Therefore $\{ C,C^\perp\}\subset\OA(\DK_{0,1})$ serve as a noncommuting common cause for the correlating events $A$ and $B$ for all possible $a_1,a_2,a_3$ values. (In fact, (\ref{ncqccs1}) fulfil for all states $\phi_\lambda$ in (\ref{failstate}) with parameters obeying  $\lambda_{AB}+\lambda_{A^{\perp}B^{\perp}}=\lambda_{AB^{\perp}}+\lambda_{A^{\perp}B}$.) The double cone support $\DK_{0,1}$ of these common cause projections is in the intersection of the backward light cones of the double cone supports $\DK_a$ and $\DK_b$ of $A$ and $B$, respectively that is in $cpast(\DK_a,\DK_b)$, which is equal to $spast(\DK_a,\DK_b)$ in this case because $\DK_a$ and $\DK_b$ are minimal double cones. (See Fig. \ref{strong}.)

Obviously, this concrete solution does not prove the validity of any of the noncommutative Common Cause Principles in the Ising model. To prove this one should find noncommutative common cause systems for \textit{any} local system.\footnote{In case of the noncommutative Weak Common Cause Principle this can be easily formulated since there is a double cone $\DK$ in the weak past of the supports of $A$ and $B$ such that the finite dimensional local algebra $\OA(\DK)$ contains $A,B$ and $C_k, k\in K$. Hence, the state restricted to $\OA(\DK)$ can be written in terms of a density matrix $\rho$ and the condition (\ref{ncqccsrew}) to be satisfied reads as
\begin{equation*}\label{ncqccsrewloc}
Tr(AB\rho_k)Tr(A^{\perp}B^{\perp}\rho_k)
=Tr(AB^{\perp}\rho_k)Tr(A^{\perp}B\rho_k),
\end{equation*}
where $\rho_k:=C_k\rho C_k, k\in K$.}
The example above only hints that abandonment of commutativity may ''compensate'' the discreteness of the model. 

Finally, we note that although the states constructed in Proposition \ref{failWCCP} are faithful they are neither space nor time translation invariant. Hence, the unitary implementation of these automorphisms in the corresponding GNS representation is not ensured although these are the only representations allowed in a decent covariant theory. Thus the status of the commutative Common Cause Principle in space and time translation invariant states is still an open question.

\section{Conclusions} 

In the paper we have shown that the validity of the Common Cause Principle in algebraic quantum field theories greatly depends on the type of the local von Neumann algebras. In Poincar\'e covariant theories they are typically of type III which, informally speaking, provide a fine enough structure to search for subprojections of arbitrary probability measure in the allowed range. In case of finite dimensional local von Neumann algebras (as in the Ising model), however, it is not obvious (and in general it is not true) that for a specific probability measure one can find commuting projections of the probability specified. But this property is crucial in the cited proof of the Weak Common Cause Principle.

One can react to this fact in two different ways. On the one hand, one may regard a discrete model as an approximation of a ''more detailed'' (discrete or continous) model. Then the failure of (commutative) Common Cause Principles can be interpreted in a similar manner to that mentioned in the Introduction: The model does not contain all the observables; hence the common cause remains burried beyond the coarse description of the physical situation in question. Only a refined extended model could reveal the hidden common causes. To make this intuition precise, one has to be explicit on what ''a refined description of the same physical situation'' means in terms of a local quantum theory given in the Introduction: $(\SDK',\OA',\cal{P}_{\SDK'},\alpha')$ is called \textit{finer} than $(\SDK,\OA,\cal{P}_{\SDK},\alpha)$ in a spacetime $\mathcal{S}$ with group $\mathcal{P}$ of geometric symmetries if there is a consistent triple $(\iota_\SDK,\iota_\OA,\iota_{\cal{P}_{\SDK}})$ of (poset, $C^*$-algebra and group) embeddings of $\SDK,\OA,\cal{P}_{\SDK}$ into $\SDK',\OA',\cal{P}_{\SDK}'$, respectively that respects also local structures, that is:
\begin{eqnarray*}
\iota_\SDK(g\cdot\DK) &=& \iota_{\cal{P}_{\SDK}}(g)\cdot\iota_\SDK(\DK),
\quad \DK\in\SDK,\ g\in\cal{P}_{\SDK},\\
\iota_\OA\circ\alpha(g) &=& \alpha'(\iota_{\cal{P}_{\SDK}}(g))\circ\iota_\OA, 
\quad g\in\cal{P}_{\SDK} ,\\
\iota_\OA(\OA(\DK)) &\subset& \OA'(\iota_\SDK(\DK)),\quad \DK\in\SDK.
\end{eqnarray*}
If the refined model is still \textit{locally finite dimensional} then the Common Cause Principle cannot be restored. Namely, the state (\ref{failstate}) can be extended to $\OA'$ which provides the same correlation without a common causal explanation due to Proposition \ref{failWCCP}. If the refined model is a kind of continuum limit (see e.g. (Vecserny\'es, Zimbor\'as, to be published) concerning the local quantum Ising model) then the weak closures of local algebras in a vacuum representation can be von Neumann algebras of type III and the Weak Common Cause Principle can be restored due to the proof of R\'edei and Summers (2002). 

On the other hand, if a discrete model is accepted as a self-contained physical model describing a specific physical phenomenon\footnote{See e.g. (Borthwick, Garibaldi, 2010) as a recent exprimental development concerning the Ising model.} then our proof shows that the \textit{commutative} Common Cause Principle cannot be a universally valid principle in local quantum theories. But why should we require commutativity for the common cause? Commutativity has a well-specified role in quantum theories: Observables should commute to be simultaneously measurable in quantum mechanics. Commutativity of observables with spacelike separated supports is one of the axioms of local quantum theory. However, as the possible causal time evolutions (\ref{causal_automorph1})-(\ref{causal_automorph2}) in the Ising model show the local observable $U_x$ does \textit{not} commute even with its own unit time translates $\beta(U_x)$ unless $\theta_1=0$ or $\pi/2$. So why not to allow common causes to be \textit{noncommuting} and see whether the Common Cause Principle can be restored in local quantum theories?
\vspace{0.2in}

\noindent
{\bf Acknowledgements.} This work has been supported by J\'anos Bolyai Research Scholarship of the Hungarian Academy of Science and by the Hungarian Scientific Research Fund, OTKA K-68195.

\section*{References} 
\begin{list}
{ }{\setlength{\itemindent}{-15pt}
\setlength{\leftmargin}{15pt}}

\item D. Borthwick and S. Garibaldi, ''Did a 1-dimensional magnet detect 248-dimensional Lie algebra?'' arXiv: 1012.540/v2 [math-ph]

\item K. Fredenhagen, ''On the modular structure of local algebras of observables'' \textit{Commun. Math. Phys.}, \textbf{97}, 79-89 (1985). 

\item R. Haag, {\it Local Quantum Physics}, (Springer Verlag, Berlin, 1992). 

\item H. Halvorson and R. Clifton, ''Generic Bell correlation between arbitrary local algebras in quantum field theory,''
\textit{J. Math. Phys.}, \textbf{41}, 1711-1717 (2000).

\item G. Hofer-Szab\'o, M. R\'edei and L. E. Szab\'o, ''On Reichenbach's Common Cause Principle and Reichenbach's notion of the common cause,'' \textit{Brit. J. Phil. Sci.}, \textbf{50}, 377-399 (1999).

\item G.J. Murphy, {\it $C^*$-algebras and Operator Theory}, (Academic Press, London, 1990). 

\item V.F. M\"uller and P. Vecserny\'es, ''The phase structure of $G$-spin models'', \textit{to be published}

\item F. Nill and K. Szlach\'anyi, ''Quantum chains of Hopf algebras with quantum double cosymmetry'' \textit{Commun. Math. Phys.}, \textbf{187} 159-200 (1997).

\item M. R\'edei, ''Logically independent von Neumann lattices,'' \textit{Int. J. Theor. Phys.}, \textbf{34}, 1711-1718 (1995a).

\item M. R\'edei, ''Logical independence in quantum logic,'' \textit{Found. Phys.}, \textbf{25}, 411-422 (1995b).

\item M. R\'edei, ''Reichenbach's Common Cause Principle and quantum field theory,'' \textit{Found. Phys.}, \textbf{27}, 1309--1321 (1997).

\item M. R\'edei, {\it Quantum Logic in Algebraic Approach}, (Kluwer Academic Publishers, Dordrecht, 1998).

\item M. R\'edei and J. S. Summers, ''Local primitive causality and the Common Cause Principle in quantum field theory,'' \textit{Found. Phys.}, \textbf{32}, 335-355 (2002).

\item M. R\'edei and J. S. Summers, ''Remarks on Causality in relativistic quantum field theory,'' \textit{Int. J. Theor. Phys.}, \textbf{44}, 1029-1039 (2005).

\item H. Reichenbach, {\it The Direction of Time}, (University of California Press, Los Angeles, 1956).

\item S. J. Summers, ''On the independence of local algebras in quantum field theory,'' \textit{Reviews in Mathematical Physics}, \textbf{2}, 201-247. (1990).

\item S. J. Summers and R. Werner, ''Maximal violation of Bell's inequalities for algebras of observables in tangent spacetime regions,'' \textit{Ann. Inst. Henri Poincar\'e -- Phys. Th\'eor.}, \textbf{49}, 215-243 (1988).

\item K. Szlach\'anyi and P. Vecserny\'es, ''Quantum symmetry and braid group statistics in $G$-spin models'' \textit{Commun. Math. Phys.}, \textbf{156}, 127-168 (1993). 

\item P. Vecserny\'es and Z. Zimbor\'as, ''Exact algebraic renormalization group and continuum limit in the Ising quantum chain'', \textit{to be published}

\end{list}

\end{document}